\documentstyle[11pt,newpasp,twoside]{article}
\def\edcomment#1{\iffalse\marginpar{\raggedright\sl#1\/}\else\relax\fi}
\reversemarginpar

\begin{document}
\title{Polarization of AGN Jets}
\author{Daniel C. Homan}
\affil{Department of Physics and Astronomy, Denison University, Granville, OH, USA}

\begin{abstract}
The sensitivity, stability, and uniformity of calibration of the VLBA has revolutionized
parsec-scale polarization studies of AGN jets.  Not only does polarization probe the 
magnetic field structures of jets, serving as a hydrodynamic tracer of shocks, bends,
and shear, but polarization also probes the medium through which it propagates by
encoding the signature of Faraday effects along the line of sight.  I review advances
made by the VLBA in studying the polarization of jets to probe their magnetic field
structures, properties of the jet plasma, and properties of the external environment.
This review covers both linear and circular polarization and is set in the context of
outstanding questions in the field.
\end{abstract}

\section{Introduction}

The Very Long Baseline Array\footnote{The VLBA is an instrument of the National
Radio Astronomy Observatory, a facility of the National Science Foundation operated
under cooperative agreement by Associated Universities, Inc.} (VLBA) has 
revolutionized the
study of parsec-scale polarization from jets of active galactic nuclei (AGN) by 
allowing highly sensitive, 
multi-frequency, multi-epoch polarization studies for a large number of sources.
In recent years, we have begun to use these capabilities to address a number of 
fundamental open questions regarding the structure, composition, environment 
and evolution of AGN jets.  

Perhaps the most glaring (and difficult) of these open questions, is that after 
forty years of study, we still do not know whether AGN flows are 
primarily an electron-proton ($e^-p^+$) plasma or a electron-positron ($e^-e^+$) 
plasma.  Another fundamental
issue is that while the synchrotron radiation we observe is produced by a 
power-law particle energy spectrum, we do not know where (or how) that 
spectrum cuts off at low energy.  These low energy particles are particularly 
important as they dominate the bulk kinetic energy of the jet (e.g. Celotti 
\& Fabian 1993).  On the fundamental question of jet
structure, the full three dimensional magnetic field configuration of jets 
is still only vaguely understood.  We know that jets contain a 
significant (and perhaps dominant) component of disordered magnetic fields 
which are shocked or sheared hydro-dynamically, but we do not know if the jets 
have an over-arching field structure (with helical, toroidal and/or 
poloidal components) which is related to the black-hole/accretion disk system.  

\section{3-D Magnetic Field Structures of Jets}

In the absence of Faraday rotation, linear polarization gives a measure of the 
net field order along the line of sight, projected onto the plane of the sky; 
however, interpretation of polarization patterns in jets is complicated by
the fact that a variety of three-dimensional magnetic field models can produce 
very similar two dimensional polarization structures.  One example is a 
helical magnetic field which will produce linear polarization that is
either parallel (Stokes $-Q$) or transverse (Stokes $+Q$) to the jet due to 
cancellation of the Stokes $U$ component from the back to the front of the jet.  
A toroidal field will also produce parallel polarization, and if viewed 
from an angle may have transverse polarization at the edges of the jet.  
Parallel or transverse polarization (as well as oblique polarization angles) 
can also be produced hydro-dynamically by shocking or shearing of magnetic
fields. 

Disentangling these effects is a difficult problem, and a first step is to
catalog the types of polarization (fractional amounts and orientations) seen
in jets.  A great deal of work has been done to catalog and compare the
parsec scale linear polarizations for a large number of AGN jets at a
variety of wavelengths (e.g. Cawthorne et al.  1993, Gabuzda et al. 2000, 
Lister \& Smith 2000, Lister 2001, Marscher et al. 2002, Pollack et al. 2003). 
Typical AGN cores are weakly polarized at the few percent level with higher
polarization being revealed at higher frequencies, and jet features typically have 
$5-10$\% polarization with a tail up to a few tens of percent.  A number of
conflicting reports have been published regarding the relative orientation of
jet polarization to jet direction at $\lambda 6$ cm (Cawthorne et al. 1993; 
Gabuzda et al. 2000; Pollack et al. 2003); averaging over these
results it seems that there is no relation for quasars, but BL Lac objects
do tend to have an excess of parallel polarization with a peak near 
$|\theta-\chi| = 0$ degrees.  Higher frequency results ($\lambda\lambda$ 1.3
\& 0.7 cm), which are less subject to contamination by Faraday rotation, show 
an excess near $0$ degrees with a broad tail for both kinds of objects
(Lister \& Smith 2000, Lister 2001, \& Marscher et al. 2002).  Marscher et al. 
(2002) have noted that this distribution is consistent with oblique shocks. 

Most parsec scale polarization observations to date are sensitive only to 
polarization in the brightest regions of jet emission.  These regions are
enhanced, most likely due to shocks or bends in the jet, and therefore the
measured polarizations may not accurately reflect the full jet magnetic
field structure.  A excellent example is the source 1055$+$018 (Attridge
et al. 1999) which with four hours of VLBA$+$Y1 time at $\lambda$ 6 cm
reveals a remarkable sheath-like polarization structure around the main
jet.  This structure was not visible in two previous one hour VLBA 
observations also at $\lambda$ 6 cm (Attridge 1999).  It is very important 
to study more sources with highly sensitive polarization observations to 
see if we are missing important pieces of the jet magnetic field structure.

Recent work by Asada (2002) and Gabuzda (2003, this volume) has suggested that 
jets may contain a significant toroidal component of magnetic field that
is visible in the form of transverse Faraday rotation measure gradients.  
Gradients in rotation measure can also form if there is a non-uniform distribution
in the external parsec-scale Faraday screen, so more work is necessary to be
sure of this interpretation.  However, the results are tantalizing, and
important three dimensional magnetic field information can be gleaned from 
Faraday rotation if it can be shown that this rotation is occurring 
internal to the jet.  Likewise, circular polarization can also yield constraints on
the three dimensional field structure. 

\subsection{A Probe of Jet Hydrodynamics: Polarization Evolution}
Observing the time evolution of jet magnetic fields is important
for understanding the magneto-hydrodynamics of jets. The VLBA has
made it possible to conduct rapid monitoring campaigns of parsec
scale jet structure.  For example, time sequences of 3C\,120 made 
by Gomez et al. (2000) at $\lambda$ 1.3 cm, reveal a wealth of detailed 
variation in the observed polarization including a possible jet-cloud 
interaction.  Interpreting these results in detail is extremely
difficult and probably requires a new generation of hydrodynamic
jet models which include magneto-hydrodynamics and full Stokes 
polarization propagation.  Such models are on the way (see Gomez, J. 2003,
this volume) and their first test will be to see if they 
can reproduce the average polarization behavior of jet components
followed in recent monitoring programs.

In Homan et al. (2002) we analyzed observations of twelve AGN jets 
from a six epoch, dual-wavelength ($\lambda\lambda$ 2 \& 1.3 cm) 
VLBA experiment to find year-long trends in polarization behavior of jet 
components over time and the fluctuations about those trends.  
Within our small sample, we found jet components tended to increase 
in fractional polarization as they moved down the jet.  We also
found there was a tendency for the polarization angles of jet 
components to rotate in the direction of being transverse to the
jet.  We argued that these two pieces of information suggested
growing longitudinal field order in the jet.  We also observed
fluctuations in polarization position angles about the longer
term trends, these fluctuations tended to be larger for more 
weakly polarized jet features as might be expected if the changes 
reflect internal changes in their magnetic field structure.
With the possible exception of 3C\,273, the polarization changes 
we observed could not be easily explained by Faraday rotation 
(or depolarization) and appeared to reflect real changes in the 
magnetic structure of jet features over time.

\section{Faraday Rotation}

Faraday rotation is the rotation of the plane of linear polarization 
during propagation of a radio wave through a magnetized plasma 
and is proportional to the square of the wavelength.  The Faraday 
rotation we observe is the net effect along our line of sight:
through our galaxy, across intergalactic space, through the
host galaxy of the quasar, and even through the jet itself.
However, it appears that Faraday rotation at centimeter 
wavelengths is dominated by screens external but near to the 
radio jet, probably in the narrow line region of the AGN
(Taylor 1998, 2000).  

Quasars typically have rotation measures of 1000 up to a few 
thousand radians/m$^2$ in their core regions and on the order 
of 100 rad/m$^2$ in their 
jets (Taylor 1998, 2000; Zavala \& Taylor 2003).  BL Lac objects
are similar to quasars with perhaps a bit lower values in
the core (Gabuzda et al. 2001,2003; Zavala \& Taylor 2003).  
Galaxies have stronger Faraday screens than either
quasars or BL Lacs and often have depolarized cores (Taylor et al.
2001; Zavala \& Taylor 2002).  Exceptions to the above rules 
do exist, the CSS quasar OQ172 was found to have 40,000 rad/m$^2$
in its core (Udomprasert et al. 1997), and Attridge et al. 
(2003, this volume) have used 86 GHz VLBA results to show that 
a rotation measure gradient of $\sim 30,000$ rad/m$^2$ must
exist between two jet components in 3C\,273.  The existence of
such a high rotation measure gradient in 3C\,273, suggests the 
need for further high frequency, high resolution Faraday rotation
measure studies of AGN cores.

Are we actually seeing narrow line clouds in these kinds of
observations?  With only a $\sim 1$\% estimated covering factor
for a uniform cloud distribution,  Zavala \& Taylor (2002)
suggest that clouds may be entrained by the jet, perhaps
in a boundary layer.  There is some direct evidence for 
jet-cloud interactions with large rotation measures observed 
in bends of 3C\,120 (Gomez et al. 2000), 0820$+$225 
(Gabuzda et al. 2001), and 0548+165 (Mantovani et al. 2002).

If the Faraday rotation does occur in a boundary layer, the
magnetic fields responsible for the rotation may be directly
related to the magnetic field structure of the jet.  Additionally,
there may be significant Faraday rotation internal to the jet
in some cases.  We argued this was the case for 3C\,279 to
explain our circular polarization observations (Wardle et al. 
1998), and Cotton et al. (2003) have seen what appears to
be internal Faraday rotation in 3C\,454.  In addition to giving us
information about the field structure of the jet, internal
Faraday rotation is highly sensitive to the low energy end
of the particle energy spectrum and combined with circular 
polarization observations may allow us to constrain that 
population. 

\section{Circular Polarization}
\label{cp}

Low resolution measurements made during the 1970s and early 1980s
found circular polarization to be only a tiny fraction
($0.1$\% was considered strongly polarized) of the integrated synchrotron 
emission from AGN jets (Weiler \& de Pater 1983, and sources therein).  
In 1988, Jones published computer simulations of the radiative 
transfer in jets showing that local levels of circular polarization
could exceed $0.5$\%, and he suggested that high resolution 
(VLBI) circular polarization observations could provide important 
constraints on jet physics.   Indeed we have found local levels
of circular polarization exceeding $0.5$\% in a handful of sources,
including 3C\,84, PKS 0528+134, PKS 0607-157, 3C\,273, and 3C\,279
(Wardle et al. 1998; Homan \& Wardle 1999, 2003).  Such strongly
polarized sources are the exception, in general we find that only
$5-10$\% of sources have local circular polarization of $0.3$\% or 
higher (Homan et al. 2001; pre-liminary results from the MOJAVE 
program, see Lister, M. 2003, this volume).  When we do detect circular
polarization, it is almost always associated with the VLBI core,
although 3C\,84 is a clear exception with well resolved structure
in circular polarization (Homan \& Wardle 1999, and ApJ submitted).

A potentially powerful diagnostic of circular polarization is the
relationship between circular and linear polarization.  In general, 
linear polarization levels exceed those for circular, but there
was no other clear relationship for a sample of 40 sources we studied
at $\lambda$ 6 cm (Homan et al. 2001).   This work needs to be repeated 
at shorter wavelengths to eliminate possible contamination by Faraday depolarization
of the linear.  Interestingly, low luminosity AGN seem to be
an exception to the above rule with circular polarization exceeding 
linear in Sgr A* (Bower et al. 1999), M81* (Brunthaler et
al. 2001), and M87 (recent MOJAVE result, unpublished data).  In our 
original study (Homan \& Wardle 1999), the two lowest luminosity objects 
in which we detected circular polarization, 3C\,84 and 3C\,273, also 
had as much or more circular polarization than linear.  This relation
between luminosity and the linear to circular polarization ratio needs 
to be investigated further; however, it may simply be due to the close 
proximity of the low luminosity objects which gives better linear
resolution.  In those nearby objects, what we are seeing as the core
is embedded in a higher density environment which could lead to external
depolarization of the linear polarization.

A key open question in the study of
circular polarization is the generation mechanism.  Circular polarization 
in AGN jets may be produced either as a direct (intrinsic) component
of synchrotron radiation or through the Faraday conversion process
which converts linear to circular polarization.  Intrinsic circular
polarization requires a significant charge imbalance in the radiating
particles (an $e^-p^+$ jet) and strong uni-directional magnetic fields
in the jet.  This combination is precisely the combination that will
lead to very large amounts of internal Faraday rotation unless the 
cutoff in the relativistic particle energy spectrum is rather high,
$\gamma_i \ga 100$ (Wardle 1977).   Faraday conversion on the other 
hand, requires a large number of low energy relativistic particles in 
the jet to do the conversion and hence a relatively low cutoff in the 
particle energy spectrum, $\gamma_i \la 20$ (Wardle et al. 1998).
A low cutoff in the particle energy spectrum may imply a jet dominated
by $e^-e^+$ pairs on kinetic luminosity grounds 
(e.g. Celotti \& Fabian 1993), and we argued this is the case for
3C\,279 (Wardle et al. 1998), but these are difficult arguments to make
conclusively.  Theoretically (e.g. Jones 1988) it appears 
easier to generate large amounts of circular polarization through
the Faraday conversion process, but direct observational evidence favoring
one mechanism over the other remains scarce (e.g. Homan \& Wardle 2003).

Another important open question is the sign consistency
of circular polarization in individual AGN.  A consistent sign of 
circular polarization would indicate a consistent underlying magnetic 
field structure, such as the poloidal field direction or the twist of 
a helical field.  Komessaroff et al. 
(1984) found a tendency for objects in their sample to show the same
sign of circular polarization over their three to five year observing
window.  We found the same to be true over a one year period for sources
undergoing strong core outbursts (Homan \& Wardle 1999).  In Homan
et al. (2001), we compared recent circular polarization results to
those from $\sim$20 years ago, and found in five of six cases the
same sign of circular polarization today as in the earlier 
measurements.  On the basis of these small statistics, we speculated 
that long-term consistency in the sign of circular polarization may be a
general property of AGN, with the sign set by the super massive 
black hole/accretion disk system.  

Bower et al. (2002) have shown convincingly that Sgr A* has indeed
maintained the same sign of circular polarization over the last 20
years.  Our time baseline for VLBA results is also growing.  Combining
results from a number of programs (Homan \& Wardle 1999; Homan, unpublished
data; M. Lister, private communication; pre-liminary MOJAVE results), there is 
now nearly seven years of VLBA data on the superluminal quasars 3C\,273 
and 3C\,279 at 15 or 22 GHz.  The sign of circular polarization in those 
objects has remained the same over that period and the signs observed today 
match those measured $\sim$20 years ago at lower frequency (Weiler \& de Pater 1983).  
Despite this progress, we still need better observational
evidence that sources in general have a preferred sign of circular polarization
and to find on what timescale any preferred sign may persist. Indeed, the fact
that there is clear evidence for changes in sign in some cases (e.g. Aller et al. 2003) 
indicates that more monitoring is needed to determine if sign changes of this type 
represent fundamental changes in overall jet magnetic field structure or stochastic 
changes in the tangled component of magnetic field. 

\section{The Future}

The future looks bright for parsec scale polarization studies of
AGN jets.  A number of fundamental questions regarding the physics of jets
and their environment remain open, and polarization can provide useful insights 
to many of them.  Over the next few years, observers will continue to exploit 
the unique capabilities of the VLBA to study polarization.  In studying the 
three-dimensional 
magnetic field structures of jets, Faraday corrected polarization maps will be 
important to remove the Faraday rotation which plagues images made at lower
frequency.  Longer integrations and greater sensitivity will also play a key
role by uncovering the full polarization structure of the jet, as in the case 
of 1055$+$018.  Faraday rotation studies and circular polarization 
observations may give important additional information about the three-dimensional 
structure of jet magnetic fields.  And following the time evolution of polarization 
will continue to give us a better handle on the magneto-hydrodynamics of jets.

In the study of Faraday screens surrounding jets, higher frequency observations will 
measure and resolve the large Faraday depths surrounding VLBI cores.  Studies of
the rotation measure distribution transverse
to jets will also be important for evaluating models of toroidal/helical
field structures.  Jet-cloud interactions seen through the effects of
Faraday rotation are particularly intriguing for what they might reveal about 
both jets and clouds.  Clear evidence for internal Faraday rotation would also 
be very important as it would help to constrain the low energy end of the particle 
distribution in jets.

Studies of circular polarization need longer time baselines to better evaluate
the issue of sign consistency in individual sources.  Better spectral studies of
circular polarization are also crucially important to constrain the emission 
mechanism; such studies will require enhanced sensitivity and improved calibration
techniques so that several frequencies can be observed quasi-simultaneously.  Higher
resolution studies will also be very useful to alleviate confusion due to the 
inhomogeneous VLBI core and to resolve the region where circular polarization is 
actually generated.


\begin{references}

\reference 
Aller, H.~D., Aller, M.~F., \& Plotkin, R.  2003,
  In: {\em Circular Polarization of Relativistic Jet Sources} eds. 
   R.P. Fender \& J.-P. Macquart,
  \apss, in press. 

\reference 
Asada, K. et al. 2002, PASJ, 54, L39

\reference
Attridge, J.~M. 1999, \apjl, 518, L87

\reference
Attridge, J.~M. 1999, Ph.D. Thesis

\reference
Bower, G.~C., Falcke, H., \& Backer, D.~C. 1999, \apjl, 523, L29 

\reference 
Bower, G.~C. et al. 2002,
\apj, 571, 843

\reference
Brunthaler, A. et al. 2001, \apjl, 560, L123

\reference
Cawthorne, T.~V. et al. 1993, \apj, 416, 519

\reference
Celotti, A. \& Fabian, A.~C. 1993, \mnras, 264, 228

\reference 
Cotton, W.~D. et al. 2003, \aap, 403, 537

\reference
Gabuzda, D.~C. \& Chernetskii, V.~A. 2003, \mnras, 339, 669

\reference
Gabuzda, D.~C., Pushkarev, A.~B., \& Cawthorne, T.~V. 2000, \mnras, 319, 1109

\reference
Gabuzda, D.~C., Pushkarev, A.~B., \& Garnich, N.~N. 2001, \mnras, 327, 1

\reference
Gomez, J.-L. et al. 2000, Science, 289, 2317

\reference 
Homan, D. C. et al. 2002, \apj, 568, 99

\reference 
Homan, D.~C., Attridge, J.~M., \& Wardle, J.~F.~C. 2001, 
\apj, 556, 113

\reference 
Homan, D.~C. \& Wardle, J.~F.~C. 1999, 
\aj, 118, 1942

\reference 
Homan, D.~C. \& Wardle, J.~F.~C. 2003,
  In: {\em Circular Polarization of Relativistic Jet Sources} eds. 
   R.P. Fender \& J.-P. Macquart,
  \apss, in press, astro-ph/0211183

\reference 
Jones, T.~W. 1988, 
\apj, 332, 678

\reference 
Komesaroff, M.~M. et al. 1984, 
\mnras, 208, 409

\reference
Lister, M.~L. 2001, \apj, 562, 208

\reference
Lister, M.~L. \& Smith, P.~S. 2000, \apj, 541, 66

\reference
Mantovani, F. et al. 2002, \aap, 389, 58

\reference
Marscher, A.~P. et al. 2002, \apj, 577, 85

\reference
Pollack, L.~K., Taylor, G.~B., \& Zavala, R.~T. 2003, \apj, 589, 733

\reference
Taylor, G.~B. 1998, \apj, 506, 637

\reference
Taylor, G.~B. 2000, \apj, 533, 95

\reference
Taylor, G.~B., Hough, D.~H., \& Venturi, T. 2001, \apj, 559, 703

\reference
Udomprasert, P.~S. et al. 1997, \apjl, 483, L9

\reference 
Wardle, J.~F.~C. et al. 1998
Nature, 395, 457

\reference
Wardle, J.~F.~C. 1977, Nature, 269, 563

\reference 
Weiler, K.~W. \& de Pater, I. 1983, 
\apjs, 52, 293

\reference
Zavala, R.~T. \& Taylor, G.~B. 2002, \apjl, 566, L9

\reference
Zavala, R.~T. \& Taylor, G.~B. 2003, \apj, 589, 126

\end{references}
\end{document}